\documentclass{iaus}
\usepackage{graphicx}

\title[Dwarf Galaxies of the Local Group] %% give here short title %%
{Dwarf Galaxies of the Local Group}

\author[Di\thinspace Stefano et al.]   %% give here short author list %%
{Rosanne Di\thinspace Stefano$^{1,2}$,   
 Roberto Soria$^1$, F.A. Primini$^{1}$, Albert Kong$^3$} 
\affiliation{$^1$ Harvard-Smithsonian Center for Astrophysics, 
60 Garden St., Cambridge, MA 02138, USA \break email: rd@cfa.harvard.edu\\[\affilskip]
$^2$Department of Physics and Astronomy, Tufts University, Medford, MA 02155\\[\affilskip]  
$^3$Center for Space Research, MIT, Cambridge, MA 02139, USA} 

\pubyear{2004}
\volume{xxx}  %% insert here IAU Symposium No.
\pagerange{119--126}
\date{?? and in revised form ??}
\jname{Proceedings Title IAU Symposium}
\editors{A.C. Editor, B.D. Editor \& C.E. Editor, eds.}
\begin{document}

\maketitle
\begin{abstract}
{\it XMM-Newton} and {\it Chandra} have ushered in a new era for the study of
dwarf galaxies in the Local Group.  We provide an overview of the
opportunities, challenges, and some early results. The large number of
background sources relative to galaxy sources is a major theme. Despite this
challenge, the identification of counterparts has been possible, providing
hints that the same mechanisms producing X-ray sources in larger galaxies are
active in dwarf galaxies. A supersoft X-ray source within $2''$ of the
supermassive black hole in M32 may be a remnant of the tidal disruption of a
giant, although other explanations cannot be ruled out.

\end{abstract}

\firstsection % if your document starts with a section,
              % remove some space above using this command.
\section{Introduction}
Dwarf galaxies constitute the largest number of galaxies residing
in groups and clusters. Dwarf galaxies also inhabit the field.
They are the building blocks of galaxy formation. Therefore,
in the young universe
and throughout vast volumes of the present-day universe, dwarf galaxies
provide the typical environments for the formation
and evolution of X-ray sources (XRSs).

A vast array of multiwavelength observations of Local Group dwarf galaxies
allow the detailed star formation histories of individual galaxies
to be derived. This is important, because theoretical considerations
and observations of larger, more distant galaxies suggest that there are
correlations between star formation history and the formation of
XRSs (see, e.g., Gilfanov 2004, Gilfanov, Grimm, \& Sunyaev 2004, and
references therein). There is, e.g., a strong correlation between star
formation rates and the formation
of high-mass X-ray binaries (HMXBs) and supernova remnants (SNRs). 
There is also a correlation between low-mass X-ray binaries (LMXBs)
and the total galaxy mass; 
the LMXB population in a large galaxy provides a measure of
the average star formation history. Detailed information
about individual dwarf galaxies  
can therefore help to both 
test and refine theory.
The Local Group dwarf galaxies in effect form a set of individual
laboratories in which the foundations of X-ray astronomy can be
studied.

We can best study X-rays from those dwarf galaxies in
the Local Group, where our observations can discover sources of 
relatively low
luminosity, and where we can determine source positions well enough
to identify counterparts at other wavelengths.
{\it Chandra} and {\it XMM-Newton} provide new tools.
Each can
resolve individual systems in the dwarf galaxies of the Local Group, where
the X-ray populations are generally not
spatially concentrated.  The large effective area of {\it XMM-Newton} 
collects enough photons to study many spectral and timing characteristics
without prohibitively long exposure times. The superb angular resolution of
{\it Chandra} allows more reliable detection of counterparts, especially
in combination with high resolution optical or radio observations;
in some cases, structures in extended sources, such as SNRs,
 may be resolved.
%Study of counterparts is particularly important for dwarf galaxies,
%because the number of sources in each galaxy is typically smaller than the
%expected background sources.

\vspace{-.3 true in} 
\section{Expected Numbers of Sources: The Background Dominates}

%\subsection{Dwarf Galaxies}

The Local Group consists of the Milky Way, M31, M33, and 
more than $30$ dwarf galaxies (see Mateo 1998; van den Bergh 2000). 
Dwarf galaxy masses range from
$< 10^8 M_\odot$ to $\sim 10^{10} M_\odot$. 
Morphological
types include dwarf spheroidals,
irregular galaxies of various shapes, and one elliptical.
Central surface densities range from $< 11.6$ to $>26$ in mag arcsec$^{-2}$,
with most values lying between $20$ and $26$.  
Some dwarf galaxies are close enough to either M31 (e.g., M32 and NGC 105)
or the Milky Way (e.g., the Sagittarius dwarf galaxy) to  
experience tidal interactions.
%Perhaps the best known of the
%Local Group dwarf galaxies are the Magellanic Clouds. 
Because of their proximity
to Earth, %and the fact that they are located well out of the Galactic plane,
the Magellanic Clouds have been well-studied
at X-ray as well as at other wavelengths. In this presentation we
will focus on the {\it other} dwarf galaxies.  These are on the frontiers
of X-ray astronomy, and present challenges that can be met by observations with the present generation of X-ray telescopes and with the next generation.    

Despite differences among the galaxies, 
we can predict the number of XRSs 
expected if the formation of XRSs
proceeds in dwarfs much as it does in larger galaxies. 
Because it is not
clear that the correlations we use are correct for masses as small as
those of individual dwarf galaxies, we make predictions only
for a set of galaxies: M32, NGC~205, IC-10, WLM, NGC~147, Sextans B,
the Sagittarius dwarf galaxy. These $7$ galaxies represent a range of
morphologies, star-formation rates, and star-formation histories.
Some are near one of the large Local Group galaxies, and some are
relatively isolated.   
In this set of galaxies we expect to find $2-10$ globular cluster XRSs;
$6-30$ LMXBs, $2-10$ HMXBs and SNRs. That is, we predict
a total 
of $10-50$ sources with X-ray luminosities near or above $10^{36}$ erg s$^{-1}$.
(Details will be presented in Di\thinspace Stefano et al.\, 2005.)    

%\subsection{Foreground and Background Sources}
\noindent {\bf Foreground and Background Sources:} 
Based on deep field studies, we expect there to be roughly $400$ XRSs per
square degree with flux ($0.5-2$ keV) greater than 
$2 \times 10^{-15}$ erg s$^{-1}$ cm$^{-2}$ (see, e.g., 
Giacconi et al.\, 2002 and references therein). One circular field of
radius $13'$, which roughly mimics {\it XMM-Newton} coverage,
should therefore contain approximately $60$ background sources.
%One square field of side $8'$, roughly mimics {\it Chandra} coverage,  
%should contain approximately 7 sources.

\begin{table}\def~{\hphantom{0}}
  \begin{center}
  \caption{Source Counts for $4$ Galaxies Observed with {\it XMM-Newton}}
  \label{tab:kd}
  \begin{tabular}{lccccc}\hline
     Galaxy & $n_{\rm H}$  & Distance & pn exp. time  
	& MOS exp. time   & Number of sources  \\
  	&   ($10^{20}$ cm$^{-2}$) & (Mpc)  & (ks) & 
	(ks) & in EPIC field\\\hline%\\%[3pt]
 NGC\,205   	& ~6.7	&   0.72   & 13.1 & 14.6 & 60\\
 Sextans B   	& ~2.7	&   1.3~   & 14.4 & 16.1 & 56\\
 NGC\,147   	& 12.0	&   0.75   & 10.8 & 12.4 & 53\\
 WLM   		& ~2.0  &   1.0~     & ~6.7 & ~9.2 & 38\\\hline
  \end{tabular}
 \end{center}
\end{table}
%\subsection{Comparisons}

\noindent {\bf Comparisons:} 
Table 1 shows the results of {\it XMM-Newton} observations for $4$ galaxies.
In each case, the total number of XRSs detected is comparable to the 
number of background sources expected. 
The impression that the background
plays the dominant role is reinforced by the fact that the
overall spatial distributions appear to be
more-or-less uniform, with little or no
clustering obviously associated with galaxy features.   
This is illustrated in Figure 1, which shows the sources detected
by {\it XMM-Newton} in the vicinity of NGC 205, a dwarf galaxy
very close to M31.   
A small fraction of the detected XRSs are associated with  
the galaxy or its environment. Note, e.g., that one of the sources in Figure 1
is associated with a globular cluster (GC).
To determine which other sources may 
also be associated with the NCG 205 or its environs, IDs with objects
at other wavelengths are crucial. The same is true for other dwarf galaxies.

\section{Individual Sources}
\vspace{-.03 true in}
\subsection{M32: Skirting a Supermassive Black Hole}
%M32 is the only elliptical galaxy in the Local Group.
%Some conjecture that it started life as a spiral, but that the arms, 
%and in fact all of the galaxy's GCs, have been tidally stripped
%by M31. 
M32 is the only Local Group dwarf galaxy
known to house a massive ($\sim 2 \times 10^6 M_\odot$) black hole (BH).
There is a clear excess of sources in the vicinity of the BH:
$3$ within a few arcseconds of the nucleus, 
while the average density is roughly $1$
per $10$ square arcminutes (Ho et al.\,2003). 
It is highly likely that the sources
near the center are there because of some activity related to the 
presence of the BH.  
Indeed, stellar interaction in the dense region
around the BH can produce XRSs.

It is especially interesting that one of the sources within $2''$
of the nucleus 
is a supersoft source (SSS). SSSs
have luminosities above $10^{36}$ erg s$^{-1}$ and emit little, if any
radiation above $1-1.5$ keV. Nine SSSs located in 
the Galaxy or in Magellanic
Clouds are associated with hot white dwarfs (WDs), and accreting WD models
provide promising explanations for the other local SSSs. But
interactions in dense stellar environments do not appear to favor the
formation of SSS WD binaries based on theoretical arguments
and on observations of GCs. Outside of GCs in M31
(which contain no SSSs),
SSSs constitute roughly $10\%$ of all observed XRSs
(Di\thinspace Stefano et al.\, 2004).   

The only other galaxy in the Local Group in which close associations
between a supermassive BH and SSSs can be detected is in M31, where
there is also a SSS within $2''$ of the central BH.
That observation inspired the conjecture the tidal stripping of
giants by a massive BH can leave behind a core which will be bright in
soft X-rays for an interval of $10^3-10^6$ years 
(Di\thinspace Stefano et al.\ 2001).
The M32 observations provide an independent venue in which to test this
conjecture and other possibilities as well. 
 
\vspace{-.1 true in}
\subsection{The Globular Cluster XRSs of NGC~205 and Sagittarius}

The X-ray luminosity function for Galactic GC XRSs 
is bimodal, consisting of ``bright" sources ($L_x > 10^{36}$ erg s$^{-1}$)
and ``dim" sources ($L_x < 10^{34}$ erg s$^{-1}$).  
The bright sources are LMXBs with neutron star accretors. Even relatively
short observations ($> 10$ ksec) with {\it XMM-Newton} or {\it Chandra} 
will detect such sources in most of the Local Group dwarf galaxies.
The dim sources are either CVs or LMXBs in quiescence. They can
only be detected in nearby dwarfs, such as the Sagittarius dwarf galaxy.
The results sketched  below indicate that the luminosity distribution of
XRSs in dwarf galaxy GCs is likely to be similar to that already studied
in the Galaxy. The discovery of additional XRSs in the GCs
of dwarf galaxies will allow more detailed comparisons to be made.   
{\bf NGC~205:} Like M32, NGC~205 is close to M31.
Soria et al. (2005) have discovered an XRS associated with one of the 
GCs, B024, in the field of the galaxy. (See Figure 1;
 Galleti et al 2004).
{\bf The Sagittarius Dwarf Galaxy}
is the nearest dwarf galaxy to us.
Ramsay \& Wu (2005) have discovered $7$ dim XRSs within the
half-mass radius of M54, one of the GCs that is near
Sagittarius and that also has a radial velocity consistent with
membership in the Sagittarius dwarf galaxy system.

\vspace{-.1 true in}
\subsection{Dwarf Galaxies with Active Star Formation}

HMXBs and X-ray active SNRs are expected only 
in galaxies which have experienced star formation within the
past $10^6-10^8$ years. 
IC 10 
and NGC 6822 both have active star
formation. 
Below we focus on two XRSs for which there is a wealth of
supporting evidence linking each to its galaxy.
{\bf NGC 6822}
contains an XRS near its center, long
suspected to be associated with a SNR. 
The combination
of {\it Chandra} observations, emission line studies by the
Local Group Survey (LGS), and radio observations have
made this identification secure. In fact, the SNR 
is resolved (see Figure 3), and the structures detected
in X-ray, radio, and optical wavelengths coincide with each other.
(See Kong et al. 2004 for details and additional references.) 
{\bf IC 10}
contains an XRS with 
$L_X(0.3-8 {\rm keV}) = 1.2 \times 10^{38}$ erg s$^{-1}$. This source
has been associated with a   
Wolf Rayet star, indicating that it is a high-mass X-ray binary.
Variations by a factor of as much as $6$ have been observed to occur
over time intervals as short as $10^4$ seconds.  The spectrum has been
fit with a multicolor disk model with $T_{in} = 1.1$ keV. (See
Wang, Whittaker, \& Williams [2005] for details; see also
Bauer \& Brandt [2004] and references therein, and Brandt et al.\ [1997].)   
These timing and spectral properties are consistent, but are not unique to
what we expect from an accreting BH. This interesting system is clearly
worthy of further study. 

\vspace{-.2 true in} 
\section{What We Have Learned, What We Can Learn}

Most XRSs found in the vicinity of Local Group dwarf galaxies
are associated with background objects. Nevertheless, previous and
ongoing studies have discovered several intriguing bright sources 
clearly associated counterparts in dwarf galaxies. Additional work 
should be able to identify galaxy 
counterparts for several times as many XRSs.
Statistical tests can also be helpful. Although they
cannot identify individual galaxy sources, they can help to quantify
the fraction of observed sources associated with the galaxies  
by quantifying the level of deviations from spatial uniformity, 
or deviations from the luminosity
function expected for the background.

The primary result derived so far is that   
the mechanisms that produce 
XRSs in dwarf galaxies appear to be
 those already well-studied in other, larger
galaxies. This is important to know if we are to estimate the
X-ray source contribution from dwarf galaxies in other parts of the
Universe.

Refined predictions
of the XRS population 
for individual dwarf galaxies, based on the star formation history
inferred from multiwavelength studies, are needed. X-ray observations can then
test the predictions, and possibly discover new effects.
One effect already discovered in both M31 and M32 is the
proximity of an SSSs to a massive BH. These SSSs could be remnants
of giants that have been tidally stripped by the BH. Additional work
is needed to determine is this is likely, or if other scenarios 
are preferred.

	Together, {\it XMM-Newton} and {\it Chandra} make a good team for
conducting studies of Local Group dwarf galaxies.
The immediate scientific returns are large. In addition,
ongoing observations can play
an important role in preparing the way for 
the next generation of   
X-ray telescopes.
The new telescopes will have significantly larger effective areas, while
spatial resolution may not be quite as good as presently available.
A modest decline in spatial resolution
will not hobble future studies of Local Group dwarf galaxies, because  
the distribution of most sources is not spatially concentrated.
On the other hand, larger effective area will provide spectral 
and timing information that can help us to better 
understand the physical nature of the sources.

\vspace{-.2 true in} 
%\cite[Lee 1971]{Lee71}; \cite{Figer02}).

\begin{center}
\begin{figure}
%\begin{center}
%\hspace{3cm}
%\scalebox{0.6}[0.6]
\scalebox{0.53}[0.53]
{ \includegraphics{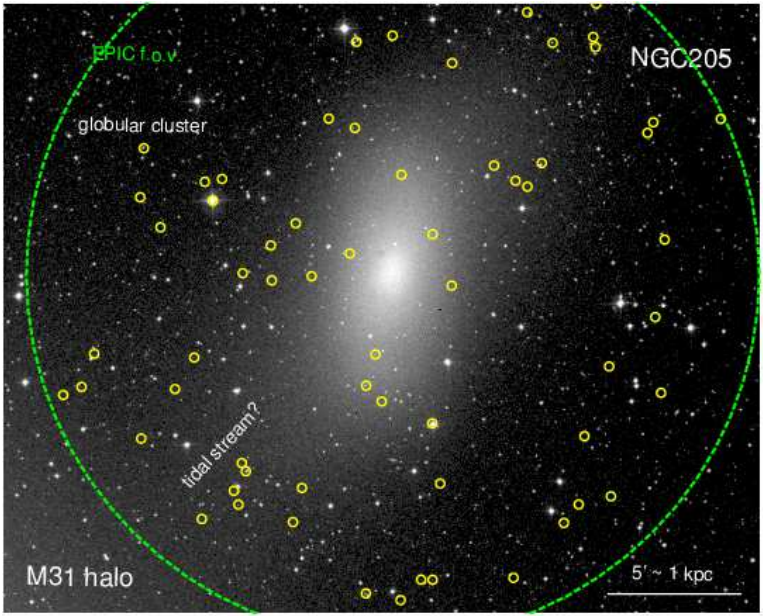}}
  \caption {XRSs (yellow open circles)
detected by {\it XMM-Newton}, overlaid on an
optical image. The position of one XRS coincides with the
position of a globular cluster.
}\label{fig1}
%\end{center}
\end{figure}
\end{center}
\begin{center}
\begin{figure}
%\begin{center}
%\hspace{3cm}
%\scalebox{0.6}[0.6]
\scalebox{0.53}[0.53]
{ \includegraphics{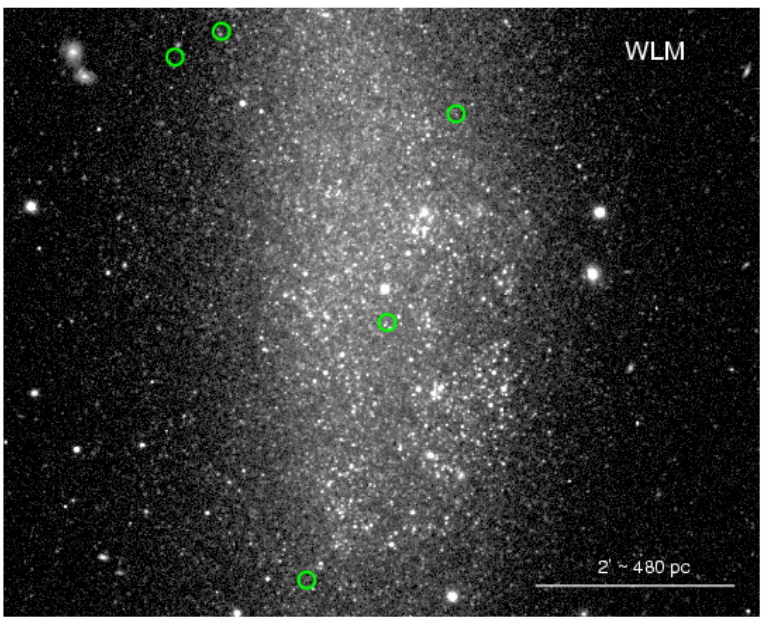}}
  \caption{
XRSs (green open circles)
detected by {\it XMM-Newton}, overlaid on an
optical image from the Local Group Survey.
}\label{fig2}  
%\end{center}
\end{figure}
\end{center}
\begin{center}
\begin{figure}
\scalebox{0.75}[0.75]{ \includegraphics{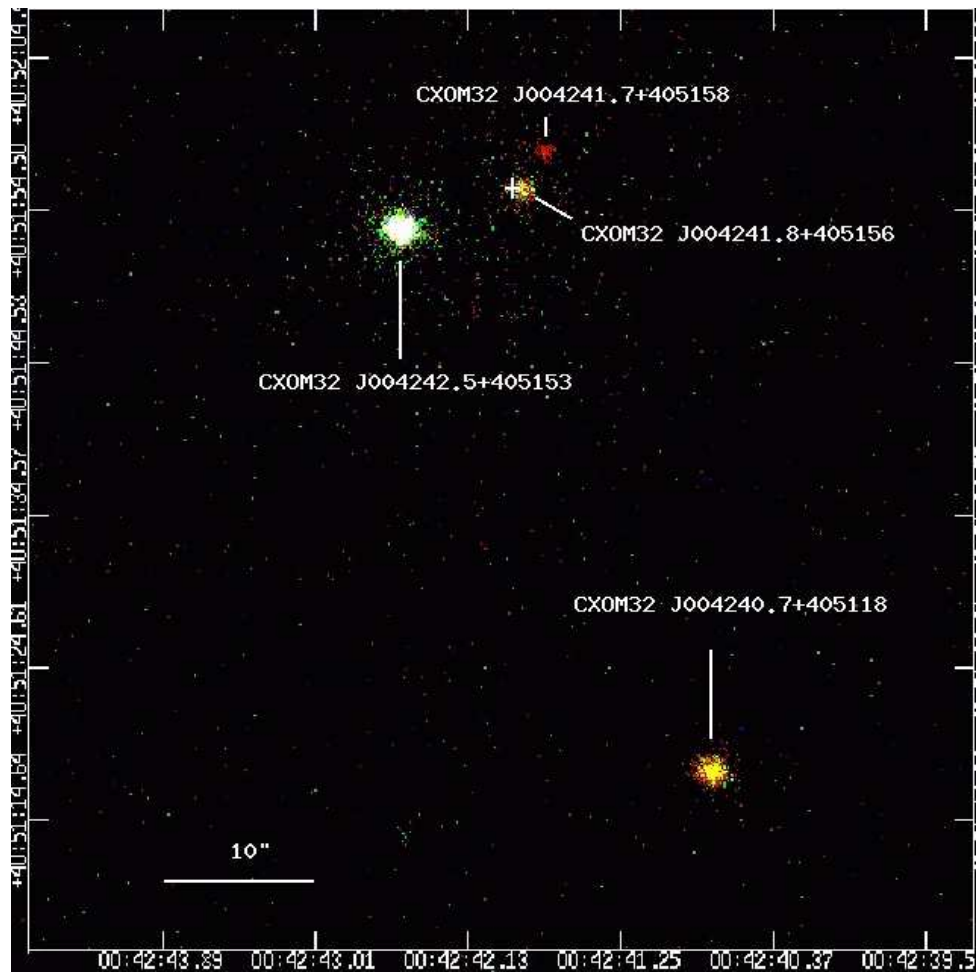}}
  \caption{M32: The white cross marks the galaxy center; the red source is
supersoft.}\label{fig3}
\end{figure}
\end{center} \begin{center} 
%\begin{center}
\begin{figure}
\scalebox{0.5}[0.5]{ \includegraphics{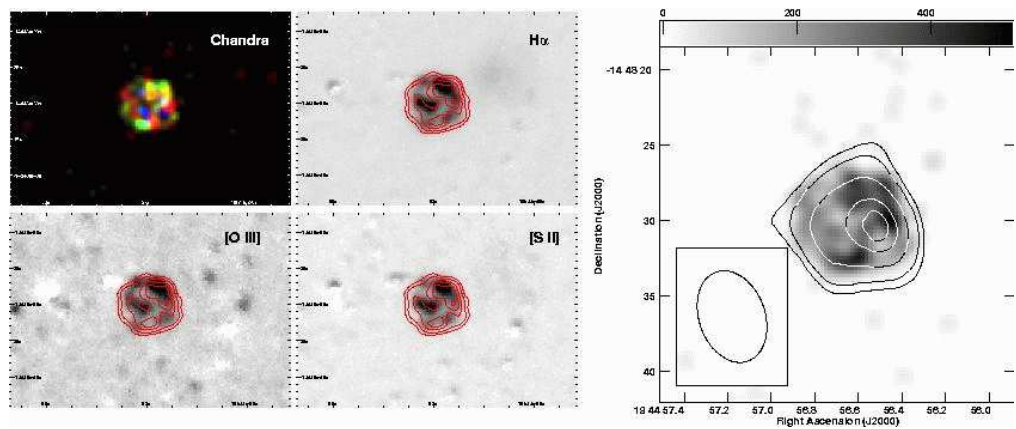}}
  \caption{SNR near the center of NGC~6822 as viewed at X-ray, optical,
and radio wavelengths.}\label{fig4}
\end{figure}
\end{center}

\end{document}